\documentclass[a4paper, 9pt]{article}

\usepackage{mathpazo}
\usepackage{url}
\usepackage{hyperref}
\usepackage{graphicx}
\usepackage{microtype}
\usepackage{caption}
\usepackage{helvet}
\usepackage{csquotes}

\usepackage[authordate, backend=biber, useprefix=false, firstinits=true]{biblatex-chicago}
\newcommand{\captiontitle}[1]{{\bfseries #1.}}
\newcommand{\subfiglabel}[1]{{\textbf{#1}}}
\newcommand{\supplref}[1]{supplement~\ref{#1}}

\title{%
    Melodic contour does not cluster:\\
    Reconsidering contour typology
}

\author{%
    Bas Cornelissen\thanks{Corresponding author: mail@bascornelissen.nl},\quad
    Willem Zuidema,\\
    John Ashley Burgoyne and
    Henkjan Honing\\[0.5em]
    \small Institute for Logic, Language and Computation,
    University of Amsterdam
}

\date{}

\begin{document}
\maketitle

\begin{abstract}
How to describe the shape of a melodic phrase?
Scholars have often relied on typologies with a small set of contour types.
We question their adequacy: we find no evidence that phrase contours cluster into discrete types, neither in German or Chinese folksongs, nor in Gregorian chant.
The test for clustering we propose applies the dist-dip test of multimodality after a \textsc{umap} dimensionality reduction.
The test correctly identifies clustering in a synthetic dataset, but not in actual phrase contours.
These results raise problems for discrete typologies.
In particular, type frequencies may be unreliable, as we see with Huron's typology.
We also show how a recent finding of four contour shapes may be an artefact of the analysis.
Our findings suggest that melodic contour is best seen as a continuous phenomenon.
\end{abstract}

\section{Introduction}
\label{sec:introduction}

Recent years have seen a renewed interest in the search for musical universals: properties common to most or even all musics across the world \parencite{Brown2011, Savage2015, Mehr2019}.
Musical universals can help to identify the constraints within which most music is made, which may, in turn, point to biological predispositions for music (\emph{musicality}) and inform theories about its evolution \parencite{Honing2018}.
The frequent use of isochronous beats is, for example, consistent with a biological, cognitive capacity for beat perception \parencite{Winkler2009}.
But music might also be shaped by physiological constraints.
A frequently cited universal is the prevalence of arch-shaped or descending melodic phrase contours, known as the \emph{melodic arch hypothesis} \parencite{Huron1996, Savage2015, Brown2011, Savage2017a}.
It has been suggested that the physiology of our vocal system explains their prevalence, making pitch contours that fall towards the end of a phrase easier to produce \parencite{Tierney2011}.

Questions of universality go hand in hand with classification: they usually require typologies that break down music into a set of \emph{characters} or \emph{features} with several possible \emph{values} or \emph{types} \parencite{Brown2011}.
Examples of features are the type of scale used or the type of rhythmic subdivision.
Both of these are \emph{discrete} characters, but there are also \emph{continuous} characters, like tempo when measured in beats per minute.
Even though it can vary almost continuously, melodic contour is often treated as a discrete character and described as \emph{ascending}, \emph{descending}, \emph{arch-shaped}, and so on.
By mapping the frequency of contour types, one can assess cross-cultural generalizations like the melodic arch hypothesis.
The validity of all such comparative questions depends on the validity of the typology used.
Consider a character \emph{modality} taking the values \emph{major}, \emph{minor}, and \emph{irregular} based on the presence of the major third.
While this seems sensible for common practice melodies, it is an awkward description of Gregorian chant or the songs of the Lakota \parencite[cf.][]{densmore1918teton}.\footnote{%
    Frances Densmore tabulated the modality of the songs she collected in precisely this way, but was well aware that this notion of modality was alien to the music she was studying.
}

In this paper, we question the validity of discrete melodic contour typologies.
Reviewing the literature on contour typologies, the common assumption seems to have been \enquote{that melodic contour types do exist and can be empirically defined} (\cite{Adams1976}, but also e.g.,~\cite{Savage2013}).
More recently, \textcite{Goldstein2025MusicPercept} present evidence for four broad common contour shapes.
We challenge this view.
Contour types do not exist, unless contours cluster accordingly.
We however fail to find any evidence for clustering in phrases from three repertoires, both within each repertoire and when aggregating them.
As a result, discrete typologies partition the contour space somewhat arbitrarily.
If the partition is not fair, it may skew the type frequencies and misrepresent the variability in the data.
All this argues for treating melodic contour as a continuous character.

\begin{figure*}[t]
    \centering
    \includegraphics[width=\textwidth]{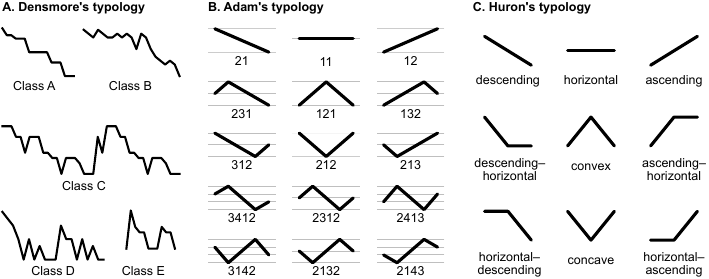}
    \caption{%
        \captiontitle{Three contour typologies}
        \subfiglabel{(A)} \textcite{densmore1918teton} divided Lakota songs into five contour classes, each identified by one representative song.
        \subfiglabel{(B)} \textcite{Adams1976} considers all possible orderings of the four boundary pitches: the type 3412 means that the initial note (i.e.,~3) is above the final (2). In between, the melody first reaches a higher (4) and then a lower (1) pitch.
        \subfiglabel{(C)} \textcite{Huron1996} considers the ordering of the average pitch on the first, middle, and final parts of a melody.
        \label{fig:contour-typologies}
    }
\end{figure*}

\section{Melodic contour typology}
\label{sec:melodic-contour-typology}

Contour is a key aspect of melody.
When still in the womb, humans already appear to be sensitive to the pitch contour of the mother tongue \parencite{Mampe2009}, and once born, contours remain a central cue for our first steps in language learning.
With such importance in early life, it is not surprising that \textcite{Dowling1978} argued that contour and scale underpin our melodic memory.
Composers who want to write catchy melodies must also attend to their contours.
Indeed, many composition treatises discuss how to shape melodies.
\textcite{Piston1970} for example opens his \emph{Counterpoint} with a chapter on the \enquote{melodic curve}, while \textcite{Perricone2018} reassures us that \enquote{there are only five basic melodic shapes or contours} (p.~179): \emph{ascending}, \emph{descending}, \emph{arch}, \emph{inverse arch}, and \emph{stationary}.
Such accounts are primarily meant prescriptively, not as a cross-cultural description of contour shapes—even though we will see some overlap.

\textcite{Adams1976} identifies a plethora of melodic contour descriptions in the academic literature.
Some narrate how the melody progresses, others settle for word lists, yet others for graphs.
Some authors propose ten types, others six, yet others four.
Descriptions are often ambiguous—how to distinguish a \emph{bow} from an \emph{arch}?—and sometimes even inconsistent.
Alan Lomax' cantometrics project, for example, coded melodic shape as \emph{arched}, \emph{undulating}, \emph{descending}, or \emph{terraced}.
But where the first three apply to the most characteristic phrase in a song, the latter applied to the entire song.
Its successor, CantoCore \parencite{Savage2012}, only includes phrase-level contour types, but six of them (\emph{horizontal}, \emph{ascending}, \emph{descending}, \emph{U-shaped}, \emph{arched} and \emph{undulating}) and the annotator is given considerable freedom to resolve ambiguities.\footnote{%
    The instructions allow coding of \enquote{clear \enquote{hyper-phrase} contours} as a single contour and advice the annotators to ignore \enquote{temporary interval changes that do not greatly affect the dominant melodic contour.}
}

An early, systematic contour analysis is Frances Densmore's \citeyear{densmore1918teton} study of the music of the Lakota people (also known as the Teton Sioux).
She visualized the contours of complete songs by plotting the accented notes (the downbeats in her transcriptions) while ignoring accidentals (\autoref{fig:contour-typologies}A).
This allowed her to cluster songs into five classes with similar contours and apparently sometimes similar social functions.
It is not entirely clear \emph{how} Densmore classified the songs.
Sometimes the global shape seems to be the crux, sometimes characteristic local features are critical, and suggest that classes were based on more than contour alone.\footnote{
    More precisely, class A is defined by its global shape, and usually has only descending intervals. Class C has a characteristic local feature (repetition of the lowest note), as does class D (an ascending opening).
}
Densmore's typology is exemplar-based: she identified one exemplary song for each class.
One could call this an \emph{inductive} typology, as it is derived from the data, also making it culture-specific.

\emph{Deductive} typologies are not culture-specific: their types are derived from first principles.
An example of this is Adams' rather intricate typology \parencite{Adams1976}.
It considers all possible orderings of a melody's four \emph{boundary pitches}: the initial note $I$, the final $F$, the first occurrence of the lowest pitch $L$, and the highest $H$.
To reformulate and simplify matters, assume that there are $k$ distinct boundary pitches, with $L=1$ the lowest and $H=k$ the highest, so that $I$ and $F$ fall in between.
Now a contour type is something like $(I=2, H=4, L=1, F=3)$ or $2\;4\;1\;3$ in short.
This means that the initial is below the final, and the melody reaches the highest and lowest pitch in between.
You can use fewer than four numbers whenever the final (or initial) is also an extreme value, as in $(I=2, L=1, H=F=3)$ or $2\;1\;3$: starting somewhere in the middle, descend to the lowest and end on the highest pitch.
All in all, there are 15 such orderings, as illustrated in \autoref{fig:contour-typologies}B.

Although Adams' paper may be the most comprehensive study of contour typology, it attracted few followers.
The typology best known today was proposed by \textcite{Huron1996} and is conceptually much simpler.
The idea is to reduce a melodic contour to three pitches: the initial $I$, final $F$, and the average pitch $M$ of all notes in between (the middle).
These pitches can be ordered in nine ways, resulting in nine possible contour types.
For example, if $I<M>F$, the contour type is \emph{convex}, if $I = M > F$, it is \emph{horizontal-descending}, and so on.
Huron also mentions a variant of the typology that divides the melody into three equal parts and uses the average pitch on the initial, middle, and final third.
This should be less sensitive to the initial and final pitch.
Following later studies \parencite[e.g.,][]{Tierney2011, Savage2017a}, we will consider this variant.

\begin{figure}[t]
    \centering
    \includegraphics[]{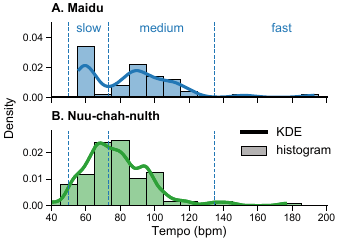}
    \caption{%
        \captiontitle{Tempo distributions of songs of the Maidu and Nuu-chah-nulth}
        Shown are histograms and kernel density estimates (\textsc{kde}).
        The tempos in Maidu music \subfiglabel{(A)} cluster in multiple groups, suggesting a typology with a \emph{slow}, \emph{medium} and possibly \emph{fast} type—which would not be appropriate for music of the Nuu-chah-nulth \subfiglabel{(B)}.
        \label{fig:tempo-typology}
    }
\end{figure}

\section{Clusterability}
\label{sec:method}

Now, what criterion should decide which typology we use?
We would argue that, as a minimum, a discrete typology should be \emph{appropriate} for the data, in the sense that the types should correspond to clusters in the data \parencite[cf.][]{Spike2020}.
Let us illustrate this using a simpler musical feature: tempo.
When measured in beats per minute, tempo is a continuous character.
A tradition might nevertheless use only a few distinct tempo ranges, such as a \emph{slow}, \emph{medium}, and \emph{fast} tempo.
If we plotted the distribution of tempos of many songs, one would expect that distribution to have three peaks or \emph{modes}.
The songs of the Maidu \parencite{densmore1958maidu} \emph{roughly} appear to follow that pattern, as illustrated in \autoref{fig:tempo-typology}A using tempo annotations from Catafolk \parencite{Cornelissen2021SysMus}.
A typology with three corresponding types (slow, medium, and fast) would therefore be appropriate for Maidu music—but it is inappropriate for the music of the Nuu-chah-nulth \parencite{densmore1939nootka}.

What we have just discussed is also known as \emph{clusterability}: the question of whether the data show signs of clustering \parencite{Adolfsson2019}.
One way to formally test this is by looking for multiple statistical modes: peaks in the probability density.
The \emph{Hartigans' dip test} \parencite{Hartigan1985} does precisely that for univariate data like the tempos.
It compares the null hypothesis that the data is unimodal with the alternative hypothesis that there are multiple modes.
The test revolves around a statistic known as the \emph{dip}: the maximal distance between the empirical cumulative distribution function and its closest unimodal approximation.
In the case of the Maidu songs, the test confirms our intuition that the tempo distribution is multimodal ($p < 0.001$), while it cannot reject unimodality for the Nuu-chah-nulth songs ($p \approx 0.08$).

The Hartigans' dip test works for univariate data but not for multivariate data like melodic contours.
A simple trick can, however, reduce the multivariate problem to a univariate one.
The \emph{dist-dip test} \parencites{Kalogeratos2012} tests whether a (multivariate) distribution is multimodal by checking whether the (univariate) distribution of pairwise distances is multimodal according to Hartigan's dip test (see \autoref{fig:synthetic}C).
After all, if a distribution is multimodal, you expect to find at least two types of pairwise distances: small within-cluster distances and larger between-cluster distances.
This means the distribution of pairwise distances is multimodal, precisely what the Hartigans' dip test can evaluate.
The dip-dist test, in short, allows us to test the presence of contour types.
We will do so using a variety of melodic data.

\section{Data and representation}
\label{sec:data}

\paragraph{Phrase contours}
We use melodic phrases from different traditions, starting with two collections of \enquote*{German} folksongs included in the Essen Folksong Collection: 1700 songs from \textcite{erk1893liederhort1, erk1893liederhort2, erk1893liederhort3} and 704 songs from \textcite{boehme1895volksthumliche}.
In addition, we analyze 152 folksongs from Nova Scotia, collected by \textcite{creighton1932nova}, and the three Chinese subsets in \emph{Essen}: \emph{Han}, \emph{Shanxi} and \emph{Natmin}.
Finally, we include phrases from Gregorian chants from the \emph{Liber Usualis} in three liturgical genres: \emph{antiphons}, \emph{alleluias}, and \emph{responsories}.
All chants come from the GregoBase Corpus \parencite{Cornelissen2020DLfM,Gregobase}, and contain breathing marks that reliably indicate phrase boundaries.
Finally, we aggregate contours sampled from each of the nine datasets into one balanced cross-cultural dataset, and we will primarily focus on that aggregate dataset.

\paragraph{Random segments}
\textcite{Cornelissen2020DLfM} report that phrases in the same chants tend to be arch-shaped, \emph{as a result of} the phrase boundaries.
If you randomly slice up the chants in segments that tend not to overlap with phrase boundaries, then the average segment has a flat contour.
To clarify whether contour types, if detected, also depend on phrase boundaries, we include the same random segments as a baseline.
Concretely, we segment each melody into random chunks by sampling the chunk lengths from a Poisson distribution fitted approximating the distribution of phrase lengths.
Omitting the first and final segment, the result is a set of segments that are roughly as long as phrases, yet unlikely to overlap with them.

\paragraph{Synthetic contours}
We also created two datasets of synthetic contours: one of which by design shows \emph{no} clustering, and one which \emph{does}.
These datasets allow us to validate the proposed methodology.
The synthetic contours are generated by a Markov process (see \autoref{fig:synthetic}A).
We sample the contour's length and initial pitch from a Poisson and binomial distribution respectively, and then walk through pitch space according to the transition probabilities observed in the cross-cultural phrase dataset.
Generating many synthetic contours in this way results in a \emph{uniform} dataset in the sense that it does not exhibit any obvious clustering structure.
By appropriately subsampling, one can create a \emph{clustered} dataset from the uniform one.
To find good cluster centers, we fit $k$-means, with $k=5$, to a dataset of 25,000 synthetic contours and then select the 1000 contours nearest to the centroids found by $k$-means.
To ensure the clusters correspond to shapes and not, say, pitch height, we used a cosine contour representation to identify neighbors \parencite{Cornelissen2021ISMIR}.
This resulted in a uniform dataset without clusters and a clustered one with five equally sized, well-separated clusters (see also \autoref{fig:synthetic}B).

\begin{figure*}[t]
    \includegraphics[width=\linewidth]{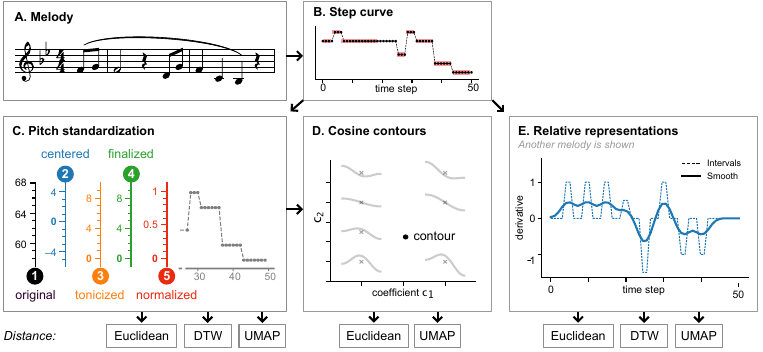}
    \caption{%
        \captiontitle{Preprocessing}
        All contours are represented as sequences of 50 pitches, sampled from a step curve interpolating the melody \subfiglabel{(A–B)}.
        Pitch is then standardized in five ways \subfiglabel{(C)}: not at all (\emph{pitch}), by \emph{centering} to mean 0, by \emph{tonicizing} or \emph{finalizing} so that the tonic or final note is 0, or by \emph{normalizing} so that the range is $[0,1]$.
        We add two relative representations \subfiglabel{(E)}: the \emph{intervals} between consecutive notes and a \emph{smooth} version thereof.
        Finally, we compute the \emph{cosine contour}, here illustrated in 2d \subfiglabel{(D)}.
        Pairwise distances are computed using Euclidean distance, \textsc{dtw} dissimilarity, or distance in a 10-dimensional \textsc{umap} projection.
        \label{fig:preprocessing}
    }
\end{figure*}

\paragraph{Contour representation}
All melodic material is processed using the same pipeline illustrated in \autoref{fig:preprocessing}.
Following earlier studies \parencite[e.g.,][]{Savage2017a,Cornelissen2021ISMIR,Cornelissen2020DLfM}, all melodic material is directly converted to fixed-length pitch sequences: we interpolate the melody with a step curve and then sample $N=50$ pitches equally spaced in time.
Using a fixed number of pitches allows us to compare phrase contours irrespective of their length.
This means we effectively normalize the phrase duration and can interpret the temporal axis as the relative position in the phrase.
Phrase length, nevertheless, has an obvious effect on contour shape: the more notes, the more shapes you can make.
To study such effects, we also record phrases' length (number of notes) and duration (in quarter notes).

\paragraph{Pitch representations}
Two further choices can obviously influence the clusterability analysis: how we represent the pitch of each contour, and how we compute distances between contours.
Besides the raw \emph{pitch} contour, we transposed the contours to make their shapes comparable irrespective of absolute pitch: we \emph{center} contours to have mean 0 \parencite[cf.][]{Savage2017a}, or transposed them so that the \emph{tonic} \parencite[cf.][]{Tierney2011} or \emph{final} note of the phrase is 0.
Next, in the \emph{normalized} version of a contour, the minimum pitch is 0, and the maximum pitch is 1 \parencite[cf.][]{Adams1976,Goldstein2025MusicPercept}.
We also include two relative representations: the \emph{intervals} between consecutive pitches, and a \emph{smoothed} version of the intervals.
Finally, we compute the \emph{cosine contour}, which describes the shape of a contour as a combination of cosine functions \parencite{Cornelissen2021ISMIR}.

\paragraph{Distance measures}
As distance measures we use Euclidean distance and \textsc{umap} distance, which is (Euclidean) distance measured in a 10-dimensional \textsc{umap} projection of the data.
Although strictly speaking no distance metric, we moreover include \emph{dynamic time warping} (\textsc{dtw}) dissimilarity.
Intuitively, if two sequences are identical except that they have warped time differently—speed up here, slow down there—their \textsc{dtw} dissimilarity is zero.
It is worth noting that a (smooth) interval representation is the derivative of the pitch contour.
In that case, the \textsc{dtw} similarity is known as \emph{derivative dynamic time warping} \parencite{Keogh2001}, which was proposed to make \textsc{dtw} more robust to small changes in the time series.
We do not use \textsc{dtw} for cosine contours.

\begin{figure*}[t]
    \centering
    \includegraphics[width=\textwidth]{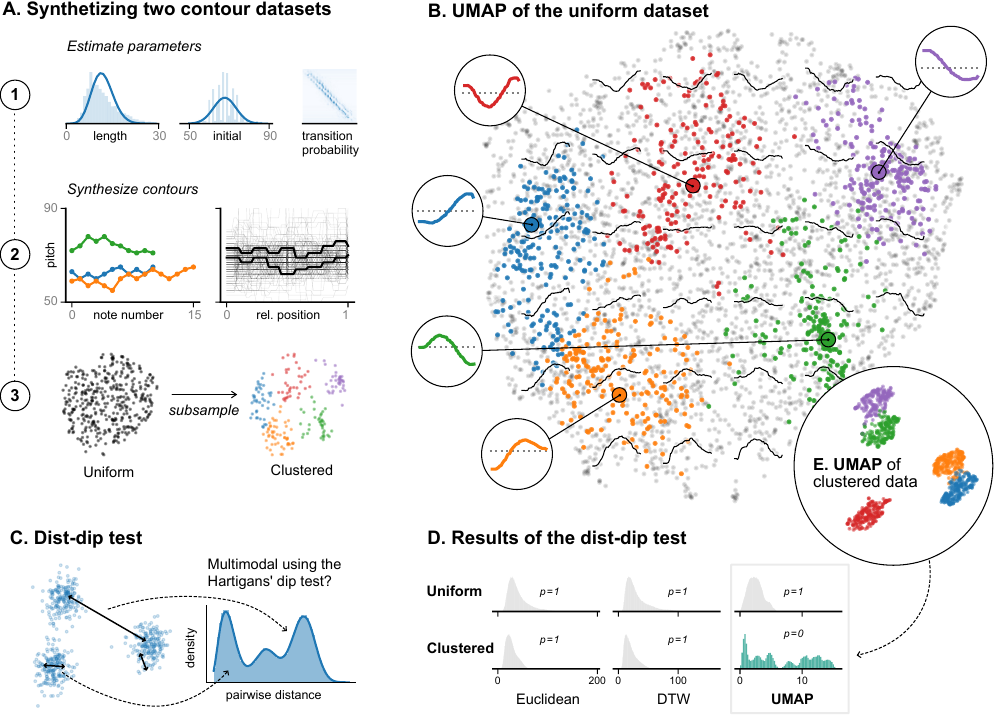}
    \caption{%
        \captiontitle{The dip-dist test discriminates clustered and unclustered synthetic datasets}
        \subfiglabel{(A)} Synthetic contours are samples from a Markov process whose parameters are estimated from actual phrase contours.
        We create a \emph{clustered} dataset by subsampling contours close to five suitably chosen cluster centers.
        \subfiglabel{(B)} A two-dimensional \textsc{umap} visualization of the \emph{uniform} dataset (gray) in which the clustered dataset (colored) is projected.
        The grid of black contours illustrates that \textsc{umap} organizes the space almost exactly like two-dimensional cosine contours (\enquote{ascendingness} horizontally, \enquote{archedness} vertically).
        The cluster centers (encircled) correspond to distinct shapes.
        \subfiglabel{(C)} The dip-dist test applies the Hartigans' dip test on pairwise distances: a multimodal dataset should have a multimodal distribution of pairwise distances.
        \subfiglabel{(D)} The dist-dip test correctly identifies the clustered dataset as multimodal but only using distances in a ten-dimensional \textsc{umap} embedding.
        This lower dimensional manifold appears more informative in that it more clearly separates the clusters \subfiglabel{(E)}.
        \label{fig:synthetic}
    }
\end{figure*}

\section{Contours do not cluster}
\label{sec:results}

\paragraph{Validation on synthetic data}
With all data in place, we turn to validating our methodology: can the dist-dip distinguish clustered from unclustered synthetic data?
Using the Python package \texttt{diptest}, a port of the R package by Martin Maechler, we apply the test to 30k pairwise distances sampled from both synthetic datasets — and find that it cannot.
The dip-dist test utterly fails to reject the null hypothesis for the clustered dataset ($p \approx 1$).
It finds no evidence for multimodality in a synthetic dataset that is multimodal by design.
Indeed, the distribution of pairwise distances seems unimodal (\autoref{fig:synthetic}D), even though the data is clustered, as is clearly visible in a low-dimensional projection made using \textsc{umap} \parencite{mcinnes2018umap}.

\textsc{umap} is a nonlinear dimensionality reduction technique that learns a low-dimensional manifold preserving the global structure of the original data.
Seeing the clusters in a \textsc{umap} projection suggests another test of multimodality: the dist-dip test using \textsc{umap} distance, which looks for multimodality in pairwise distances between contours measured in a lower, ten-dimensional manifold.
Indeed, this test \emph{does} correctly reject the null hypothesis for the clustered dataset, but not for the uniform one (\autoref{fig:synthetic}D).
\textsc{umap} distances apparently capture the cluster structure of synthetic contours better than Euclidean distance does.

One may question using projected data to test for multimodality, since the projection can now heavily influence the result.
Dimensionality reduction techniques like \textsc{umap} can sometimes suggest clusters that are not present in the data, possibly making the multimodality test overly sensitive.
Importantly, this would only strengthen a negative result: if we do \emph{not} even find evidence for multimodality using \textsc{umap} distances, clusters are only more likely to be absent.
Note also that the use of a projection is directly comparable to how principal component analysis is sometimes used before statistical testing in other clusterability approaches \parencite{Adolfsson2019}.
Moreover, one can think of all this as a more rigorous alternative to visually inspecting low-dimensional visualizations for signs of clustering.

\begin{figure*}[!t]
    \centering
    \includegraphics[width=\textwidth]{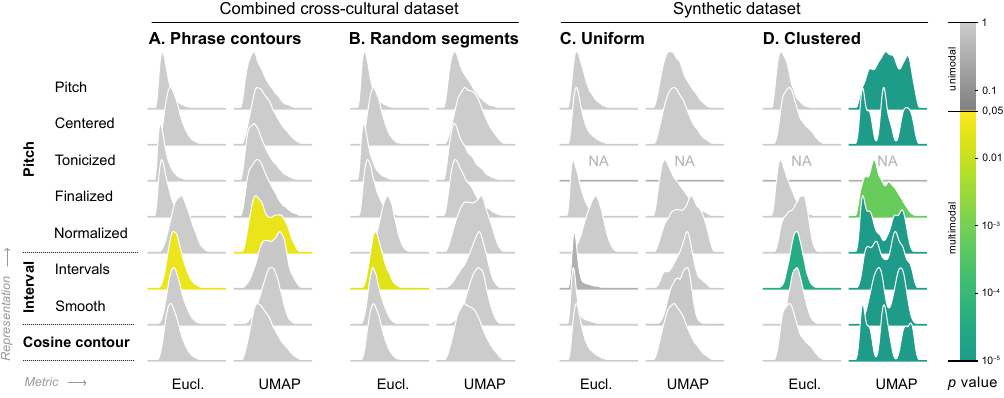}
    \caption{%
        \captiontitle{Melodic phrase contours do not cluster}
        Shown are the distributions of pairwise distances between contours in various conditions.
        If contours cluster, we expect multimodal distance distributions.
        We test this using the Hartigans' dip test and let colors indicate $p$-values, such that grey distributions are not significantly multimodal ($\alpha=0.05)$.
        Eight different representations (vertically) and two metrics (horizontally) are analyzed: Euclidean distance and the distance in a lower-dimensional \textsc{umap} embedding.
        The latter successfully discriminates unclustered from clustered synthetic data (\subfiglabel{C} vs.~\subfiglabel{D}; see also \autoref{fig:synthetic}).
        However, neither in phrases \subfiglabel{(A)} nor in random segments \subfiglabel{(B)} does the test find clear evidence for clustering.
        \label{fig:results}
    }
\end{figure*}

\paragraph{Main results}
Equipped with a method that can detect clusters in synthetic contour data, we turn to the actual phrase contours.
\autoref{fig:results} shows the distribution of pairwise distances for phrase contours and random segments, along with the two synthetic datasets.
The color coding indicates $p$-values and highlights that the dist-dip test on \textsc{umap} distances only rejects unimodality for the clustered, synthetic dataset—not for any of the phrase contours.
In other words: melodic contours do not appear to cluster.

To rule out that this is an artifact of the representation, we repeat the analysis with the eight different representations shown in \autoref{fig:preprocessing}.
With none of the eight representations do we find evidence for clustering in either phrase contours or random segments, using any of three similarity metrics: Euclidean, \textsc{dtw}, and \textsc{umap} distance.
The only possible exception is the interval representation.
With that representation, however, the uniform synthetic contours also appear to be clustered, even though they are designed not to be.
The conclusion does not change if we only consider unique contours, reduce the dimensionality of the contours from 50 to 10, or analyze individual datasets separately (see \supplref{suppl:clusterability}).

One may expect the length of contours to have an effect: there are simply fewer possible shapes when you have only four notes instead of ten, and so you should see more clusters amongst shorter phrases.
If we split out our analysis by length, the dist-dip test on \textsc{umap} distances indeed indicates multiple modes for the smaller phrases up to 5 notes, and sometimes also for the longest ones of around 15 notes or more.
But for the most common phrases, with average lengths between 5 and 15 notes, we find no convincing evidence for clustering of phrase contours.
In contrast, we do find such evidence for the synthetically clustered dataset (see \supplref{suppl:length-wise-analysis}).
In sum, our method produces no evidence of clustering of melodic contours.

\section{Implications}
\label{sec:implications}

Our findings directly challenge the idea that \enquote{contour types do exist and can be empirically defined} \parencite{Adams1976}, at least for melodic phrases.
We would like to discuss the implications for discrete typologies.

\begin{figure*}[t]
    \centering
    \includegraphics[width=\textwidth]{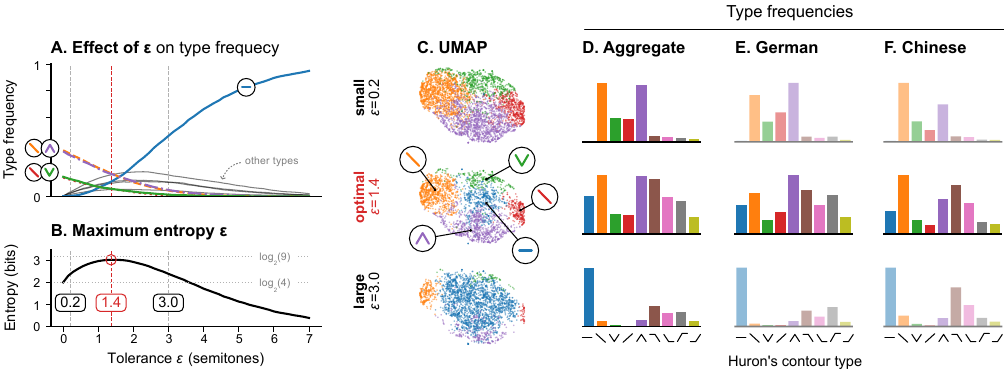}
    \caption{%
        \captiontitle{A tolerance parameter $\epsilon$ distorts type frequencies in Huron's typology}
        This typology compares the average pitch on a contour's start, middle, and end, considering two pitches equivalent if their absolute difference is below a tolerance $\epsilon$.
        The value of $\epsilon$ strongly influences the type distribution \subfiglabel{(A)}.
        We propose to choose the $\epsilon$ that divides contours over types as evenly as possible \subfiglabel{(B)}.
        While a small value ($\epsilon=0.2$) obscures partly horizontal types, a large one ($\epsilon=3.0$) exaggerates it \subfiglabel{(C)}.
        Shown is a \textsc{umap} projection of aggregate phrase contours for only 5 out of the 9 types.
        Frequencies of all types are shown as histograms \subfiglabel{(D-F)}.
        \label{fig:tolerance}
    }
\end{figure*}

\paragraph{Skewed type frequencies}

The most important implication is the following.
Every discrete typology partitions the contour space in different types.
If the data consists of clearly separated clusters, like the clustered synthetic data, the data suggest an obvious way to partition the space: one cluster per type.
In that case, one can also reasonably compare type frequencies.
In the absence of clustering, however, \emph{any} partition of the space will be somewhat arbitrary, and type frequency are easily misunderstood.
This is neatly illustrated by Huron's typology, where a hidden parameter turns out to strongly affect type frequencies.

Recall that \textcite{Huron1996} compares the average pitch over three segments of a melody.
These are rarely exactly identical and so he, presumably, considers pitches as equivalent if their absolute difference is below a tolerance threshold $\epsilon$.
With zero tolerance ($\epsilon=0$ semitones) horizontal contours will be extremely unlikely, but with a tolerance of an octave ($\epsilon=12$ semitones) virtually any contour will be considered horizontal.
And so this tolerance parameter can strongly distort the type distribution,  
as illustrated in \autoref{fig:tolerance}.
\textcite{Huron1996} indeed reports four highly frequent types (ascending, descending, convex and concave), while the remaining five types are very rare (each $\le1.2\%$ of the data).
This is exactly the expected pattern for a small tolerance parameter (\autoref{fig:tolerance}E, first row).
Importantly, the primary reason these five types are infrequent is mathematical (i.e., the choice of $\epsilon$)—not musical.
It only suggests horizontal contours are rare because the type definition is biased to make them rare.

Since Huron's typology is commonly used, what would be a good choice of the tolerance parameter $\epsilon$?
\textcite{Tierney2011} use $\epsilon=0.2$ semitones without motivation, and Huron does not report a choice of $\epsilon$ at all.
Our argument suggests choosing $\epsilon$ so that it makes the type distribution as even as possible.
This would mean choosing $\epsilon$ to maximize the entropy of the type distribution.
When estimating $\epsilon$ on the \emph{cross-cultural} phrase contour dataset, we find that $\epsilon=1.4$ semitones maximizes the entropy (\autoref{fig:tolerance}B).
This is much higher than the values previously used and suggests that those studies may indeed have misrepresented type frequencies.
Note that all this carries over to Adams' typology, which also relies on a tolerance parameter.

\begin{figure*}[t]
    \centering
    \includegraphics[width=\textwidth]{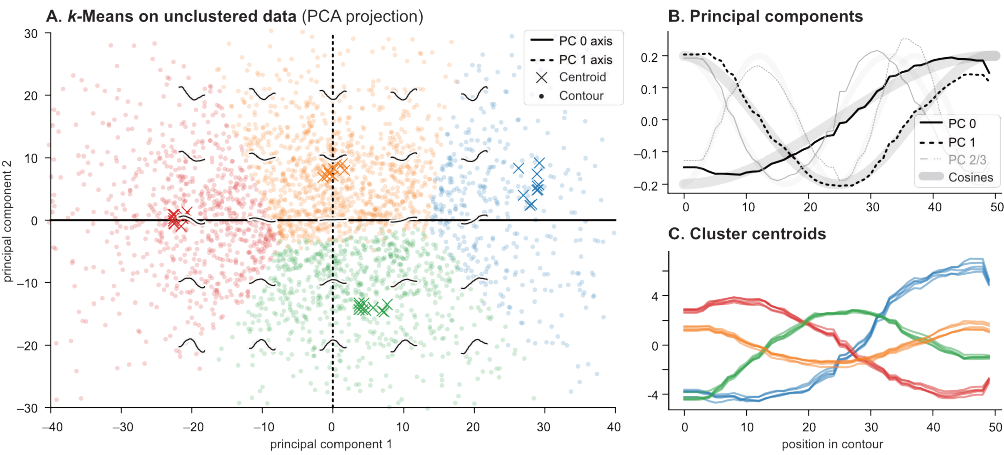}
    \caption{%
        \captiontitle{k-Means centroids reflect cosines}
        A 2d principal component projection of unclustered synthetic contours is shown, colored according to a $(k=4)$-means clustering \subfiglabel{A}.
        A grid of black contours in \subfiglabel{(A)} illustrates how the space is organized, with the first two components essentially measuring ascendingness and concavity.
        More precisely, the principal components of such data approximate cosines of increasing frequency \subfiglabel{(B)}.
        The clustering is stable across bootstrapped runs, as shown by the close centroids (crosses in A), and by their similar shapes \subfiglabel{(C)}.
        The centroids consistently fall close to the principal axes, resulting in ascending, descending, convex and concave shapes---for primarily mathematical, not musical reasons.
        \label{fig:k-means}
    }
\end{figure*}

\paragraph{Spurious shapes}
Alternatively, one can turn to an \emph{inductive} typology: learn a typology fitting for the data, in the spirit of \textcite{densmore1918teton}.
\textcite{Goldstein2025MusicPercept} take exactly this approach when they use $k$-means to cluster contours from the Essen folksong collection.
They report evidence for \enquote{presence of four broad common contour shapes} \parencite{Goldstein2025MusicPercept}: ascending, descending, convex, and concave.

To understand this seemingly contradicting result, note that \textcite{Goldstein2025MusicPercept} themselves report that neither the elbow method nor silhouette scores convincingly indicated a choice of the number of clusters.
This is consistent with our finding that contours do not cluster.
And so they base their choice for $k=4$ clusters entirely on the type imbalance observed by \textcite{Huron1996}: 4 very common types, 5 very rare ones.
Critically, as we have seen, Huron's observation is an artefact of his choice of tolerance parameter. 
It primarily reflects a property of his analysis, not of the music.
Nevertheless, one may argue, it is striking that the four clusters learned by $k$-means are entirely consistent with Huron's four most common types. 
Consistent types in two (almost) independent typologies is compelling evidence—except when it is another artefact, in this case of the principal component analysis.

In short, \textcite{Cornelissen2021ISMIR} show that temporally correlated sequences like melodies tend to have principal components shaped like cosine functions of increasing frequency, as illustrated in \autoref{fig:k-means}.
\textcite{Goldstein2025MusicPercept} indeed apply a principal component analysis to their centered contours.
Focusing on the first two components, the first principal axis will order shapes from descending to ascending, and the second axes will order shapes from convex to concave (the order may be flipped, but that is inconsequential).
This pattern is illustrated by the grid of black contours in \autoref{fig:k-means}A.
Even if dimensionality of the data is much bigger, in the absence of clustering, $k$-means tends to place its four centroids close to the first two principal axes, which indeed typically explain most of the variance \parencite{Cornelissen2021ISMIR}.
The resulting centroids are ascending, descending, convex and concave: again for entirely mathematical, not musical reasons. 
If that did not convince you that the effect should not be interpreted musically: \autoref{fig:k-means} used \emph{synthetic} contours all along.

In summary, we find no evidence for clustering, let alone four broad contour types.
Still, $k$-means will happily divide contours in an arbitrary number of clusters.
With $k=4$ it happens to produce clusters for familiar shapes, but for entirely mathematical reasons. 
While the overall cluster shape of $k=4$ centroids may be an artefact, their deviations from perfect cosines are still informative.
For example, all contours start and end flat because every melody is necessarily stable for the duration of the first and final note.
At least $k$-means reflects such tendencies, while deductive typologies do not.

\begin{figure*}[t]
    \centering
    \includegraphics[width=\textwidth]{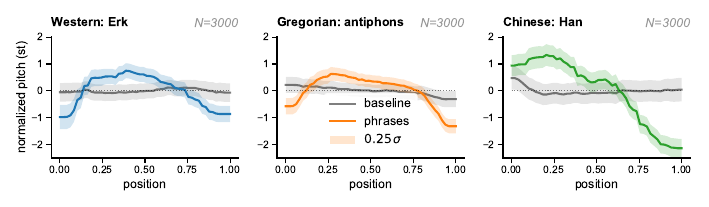}
    \caption{%
        \captiontitle{Average phrase contours differ across three traditions}
        The average phrase contours of German folksongs, Gregorian chant, and Chinese folksongs compared to baselines of random melodic segments from the respective corpus (gray).
        They support the overall tendency for arch- \emph{or} descending average contours but show interesting differences: the Chinese average is \emph{not} arch-shaped and, strictly speaking, a counter-example to the melodic arch hypothesis.
        This illustrates how a continuous approach to contour typology can directly reveal cultural differences.
        \label{fig:average-phrase-contours}
    }
\end{figure*}

\paragraph{Continuous typology}
All in all, the key implication seems to be that contour is a continuous phenomenon. 
Discrete descriptions, unless used carefully, can be misleading.
And in fact, discrete typologies may not even be necessary, as \textcite{Huron1996} already anticipated. 
We can directly compare the average phrase contours of different traditions, as shown in \autoref{fig:average-phrase-contours}.
The gray lines in that figure show the average shapes of random segments, implying that the shape of the average phrase contour results from the particular placement of the phrase boundaries.
The average contours also reflect tendencies present throughout the data, such as the initial and final flattening noted above.

This analysis also allows one to compare contour corpora.
While the German and chant phrases tend to be arch-like, the average phrase contour of Chinese folksongs looks quite different: not only is the range much larger, its shape is best described as descending, or perhaps horizontal–descending.
This is consistent with the type differences between German and Chinese songs found using the maximum-entropy version of Hurons typology (\autoref{fig:tolerance}E vs.~F, middle row).
One may even go so far as to take this as a counter-example to the strict formulation of the melodic arch hypothesis, namely, that melodies tend to have arch-like shapes, which brings us back to the questions motivating all this work.


\section{Conclusions}
\label{sec:conclusions}

Melodic contour does not cluster.
Approaching this as a clusterability problem, we analyze phrase contours from three musical traditions, yet find no evidence that contours form discrete types.
This is unlikely to be a result of methodological choices, since we repeat the analysis with a wide variety of representations and distance measures.
Moreover, the clusterability method can successfully distinguish between clustered and non-clustered synthetic data.
We explain how earlier studies could reach seemingly contradictory results and discuss implications for discrete contour typologies, notably how they can skew type frequencies.
Although we propose remedies, the deeper implication is that melodic contour is a continuous character.

A shortcoming of this work is the limited cross-cultural validity of the data analyzed.
Except for \textcite{Savage2017a}, most previous studies have relied on the Essen Folksong Collection, and this study only added Gregorian chant as a third tradition.
However, our central finding—that contour shapes do not cluster—is negative.
For that, cross-cultural validity is not as much of an issue: even the limited data we analyzed serves as a counter-example.
The same is true when rejecting two formulations of the melodic arch hypothesis.
But we think that the methods we proposed, and the continuous methodology we argued for, are sufficiently general to be applicable in other traditions—or even different domains.

Phrase contours, after all, are not only studied in music but also in language.
The study of intonation in phonology has produced various cross-cultural generalizations, such as the \emph{decline} from the beginning towards the end of a phrase, or the start of a phrase by a sharp rise known as the \emph{reset} \parencite{Ladd2001}.
At the same time, models have been proposed to describe the intonation contours found in particular languages, such as the ToBI system in English \parencite{Silverman1992}.
This revolves around a grammar for combinations of high and low tones and gives rise to a similar set of questions addressed in the present paper.
One recent study, for example, used functional data analysis (\textsc{fda}) to analyze the pitch contours of falling and rising intonation types in English.\footnote{%
    This seems comparable to using a cosine contour representation, assuming that cosines indeed approximate the principal components, as is the case for melodic contours.%
}
Although the authors do not explicitly test for this, as we do here, the results suggest that these contours form clusters \parencite{Zellers2010}.
Analogous to this paper, the authors move from a discrete analysis (ToBI) of intonation contours to a continuous one (\textsc{fda}).
A more recent study \textcite{Gerazov2021a} uses $t$-\textsc{sne} to visualize intonation contours, and an obvious next step would be to apply the clusterability methods developed in this paper to verify whether those contours indeed cluster.
More generally, this convergence calls for an interdisciplinary study of contour in speech and song.

\section*{Data and code}
Data and code have been released via \href{https://github.com/bacor/shapes-of-music}{github.com/bacor/shapes-of-music}.


\printbibliography

\clearpage
\appendix
\renewcommand\thesection{S\arabic{section}}
\setcounter{figure}{0}
\renewcommand{\thefigure}{S\arabic{figure}}
\renewcommand{\thetable}{S\arabic{table}}
\renewcommand{\theequation}{S\arabic{equation}}

\noindent{\LARGE\bfseries Supplementary Materials}
\medskip

\section{Clusterability of contours}
\label{suppl:clusterability}

Here we show the $p$-values of the Hartigans' dip test on the set of pairwise distances between contours, using Euclidean, \textsc{dtw} and \textsc{umap} distance.
The color coding is the same as in the main text: yellow-green for significant results, gray for insignificant results, using a significance threshold of 0.05.
Only with \textsc{umap} distance does the test correctly provide evidence for multimodality in the clustered dataset \subfiglabel{(D)}.
Note also that the interval representation finds more evidence for multimodality across all four datasets—even in the uniform, synthetic dataset \subfiglabel{(C)}.
But since that dataset is synthesized to contain \emph{no} clusters, we treat this as a false positive.

\begin{figure*}[!h]
\centering
\includegraphics[width=\linewidth]{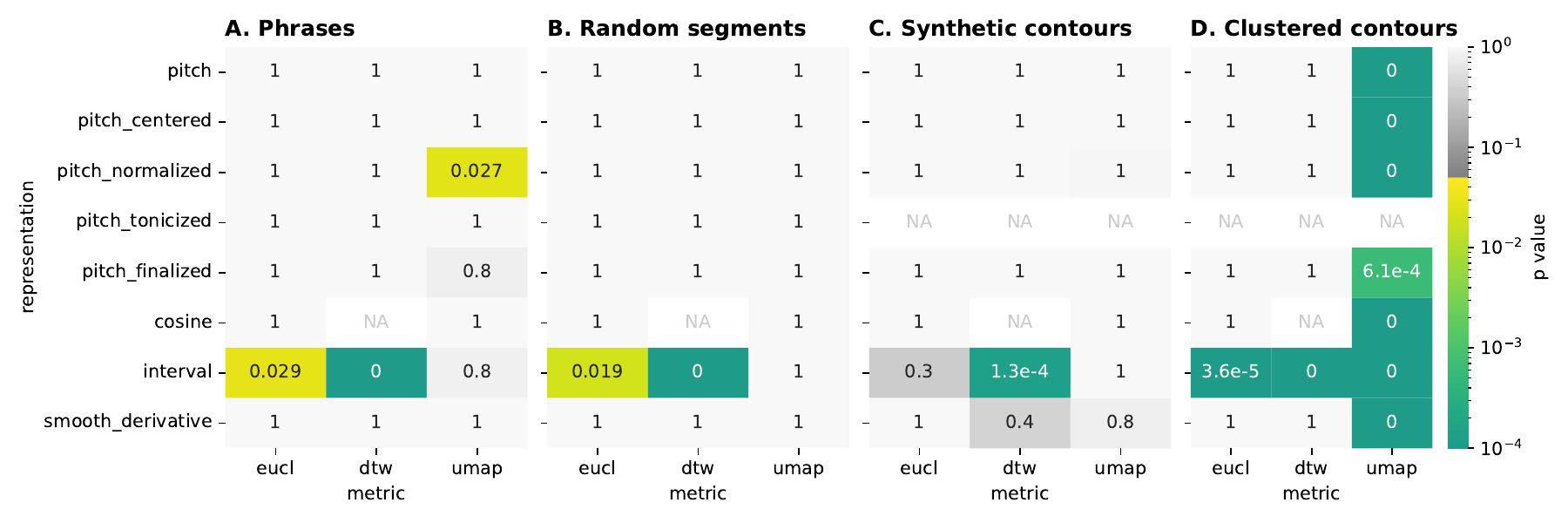}
\end{figure*}

\paragraph{Unique contours only}
We repeated the analyses on samples of unique contours, and the overall pattern remains the same.

\begin{figure*}[!h]
\centering
\includegraphics[width=\linewidth]{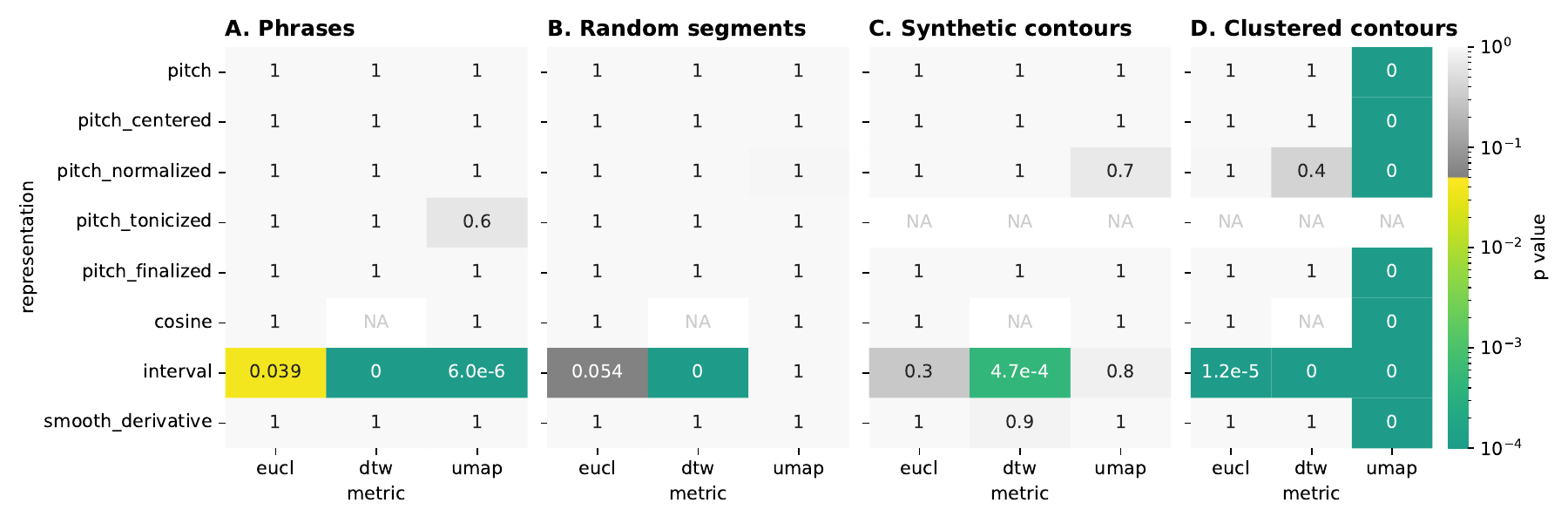}
\end{figure*}

\clearpage

\paragraph{Lower dimensionality}
We repeated the analyses on lower-dimensional contour representations of 10 rather than 50 pitches, by subsampling the 50-dimensional contours, or in the case of cosine contours, taking only the first 10 coefficients.
Again, the overall pattern remains the same:

\begin{figure*}[!h]
\centering
\includegraphics[width=\linewidth]{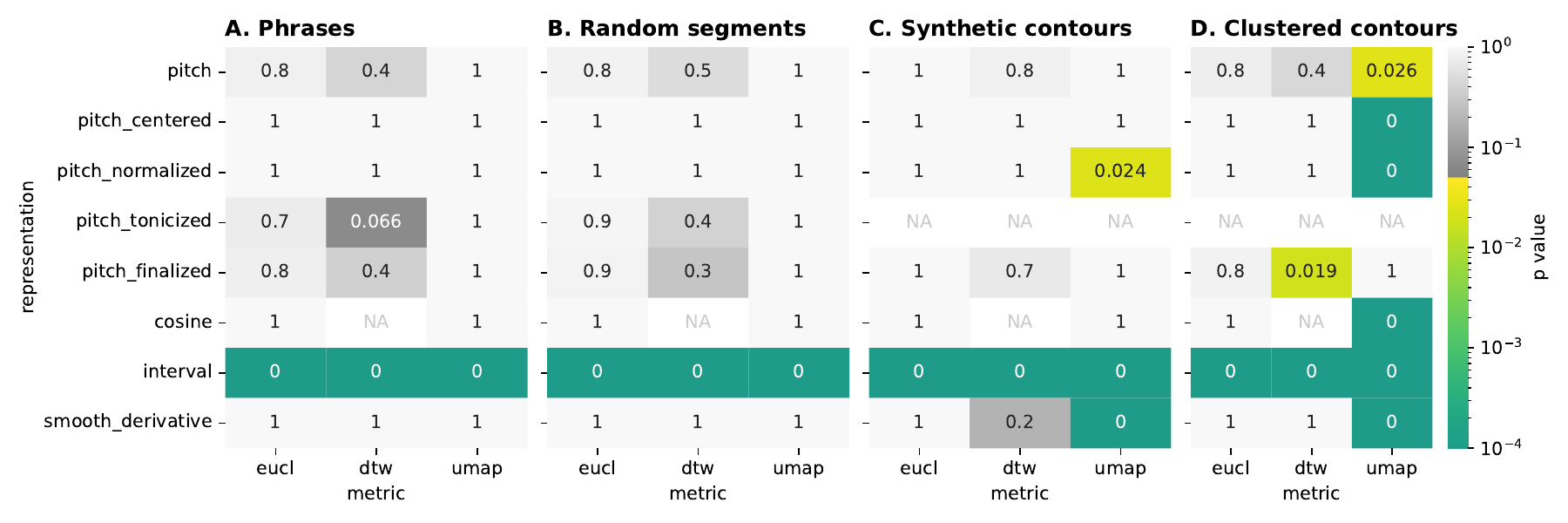}
\end{figure*}

\paragraph{Per dataset}
The results for `phrases' above all use the aggregate, cross-cultural dataset.
Here we show the results for three datasets separately.
Again, the overall pattern remains the same.

\noindent
\includegraphics[width=.99\textwidth]{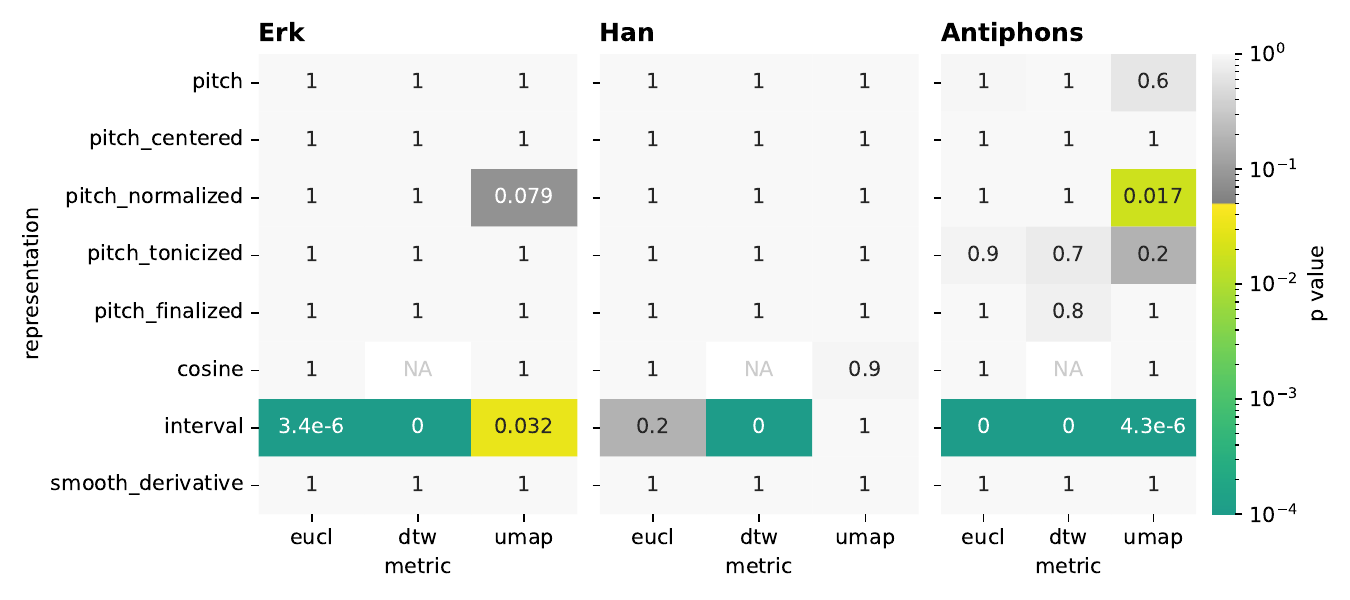}

\clearpage

\section{Length-wise analysis}
\label{suppl:length-wise-analysis}

The length of a phrase may affect its shape, and perhaps we don't find clusters because we aggregate all lengths.
We thus repeat the analyses, but now for every length (measured in the number of notes) separately.
First, this is the distribution of lengths in the datasets:

\bigbreak\noindent\includegraphics[width=.7\textwidth]{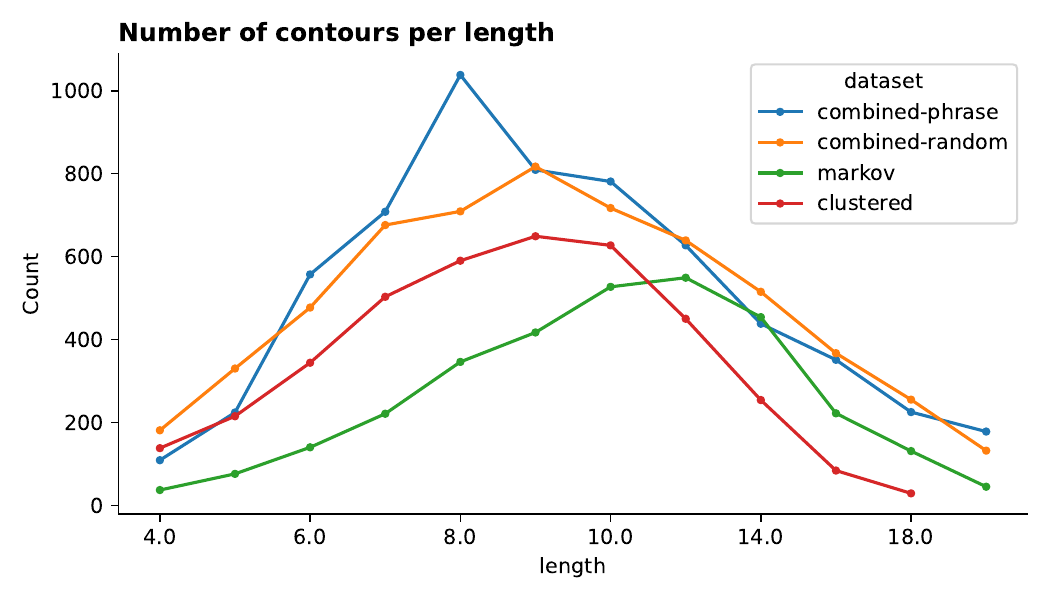}

\paragraph{Euclidean distance}
Next, we show the same $p$-values as before, but now the length is shown vertically, and the representation horizontally.
With Euclidean distance, we only see evidence appearing for some clusters of very short contours of 4--5 notes.
This is not surprising: the space of possibilities is small, and there are only a few such contours.
Indeed, even uniform synthetic contours of length four can appear clustered.
Note that many of the synthetically clustered contours still avoid detection.

\begin{figure*}[!h]
\centering
\includegraphics[width=.99\linewidth]{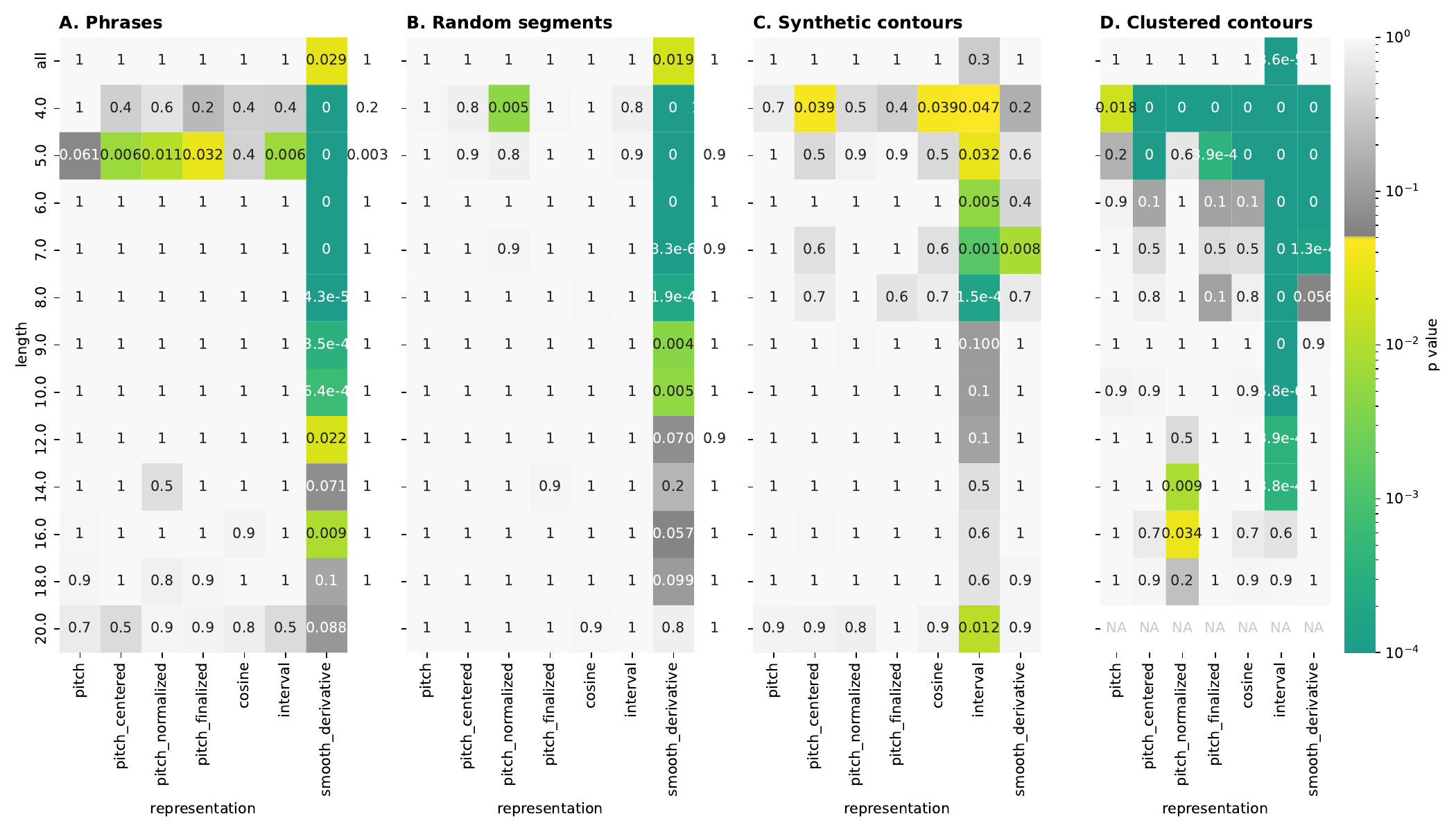}
\end{figure*}

\pagebreak

\paragraph{\textsc{umap} distance}
With \textsc{umap} distance, we see more evidence for clustering, but again primarily for shorter motifs, and in particular with the normalized pitch representation.
There is also some clustering for longer phrases.
But for the most common phrases of average length, that evidence is largely absent and certainly not nearly as strong as the evidence for clustering in the synthetic, clustered dataset.

\begin{figure*}[!h]
\centering
\includegraphics[width=.99\linewidth]{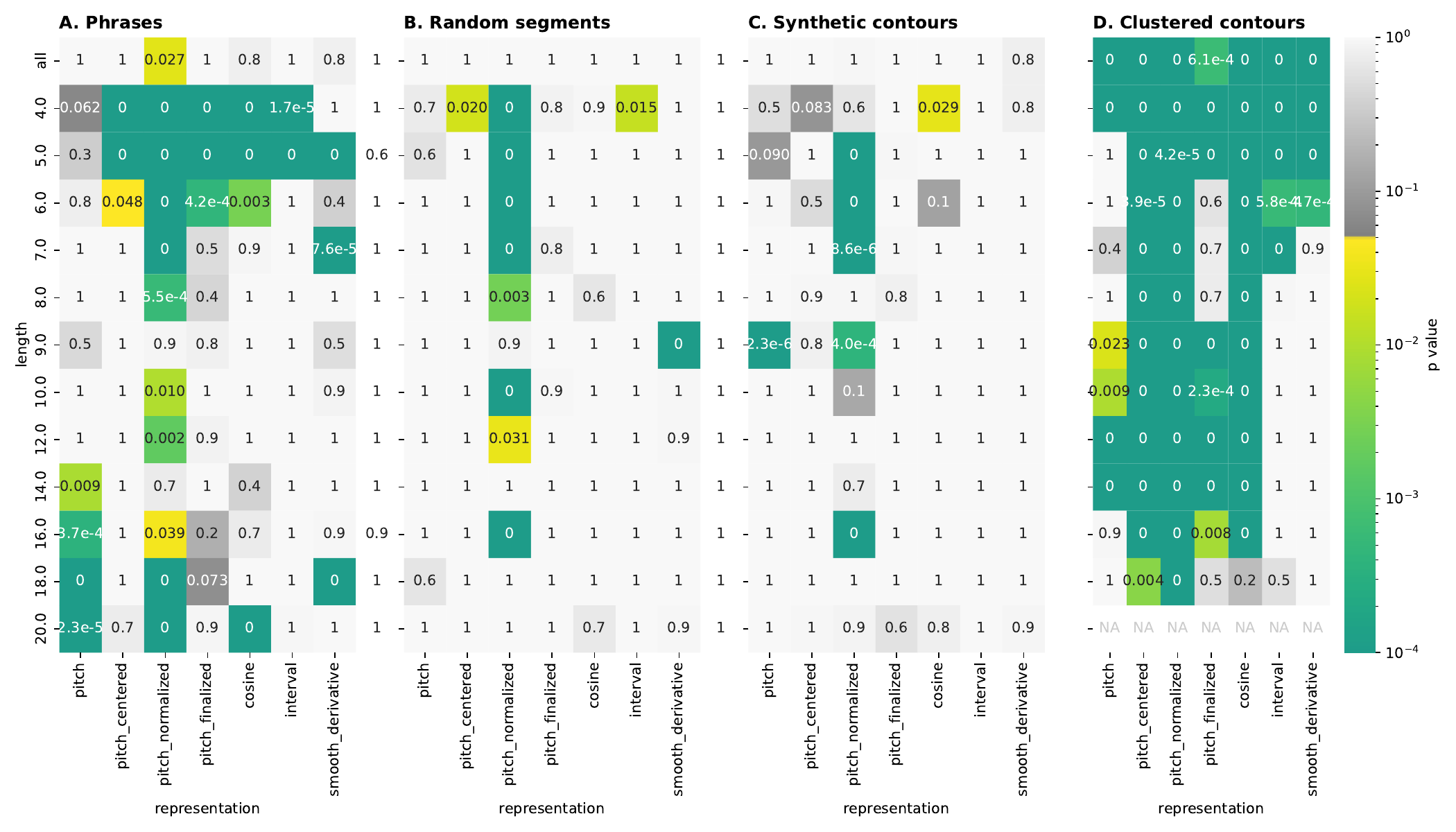}
\end{figure*}

\clearpage

\section{Average shapes}
\label{suppl:average-shapes}

\begin{figure*}[!h]
    \centering
    \includegraphics[width=.99\linewidth]{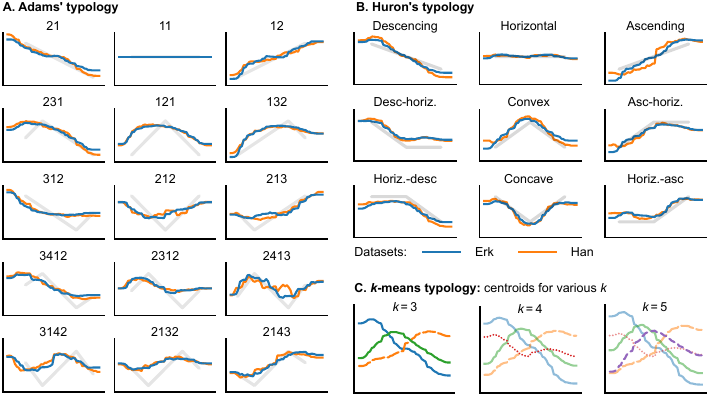}
\end{figure*}

\noindent%
We show the average of all contours with a certain type for Adams' \subfiglabel{(A)} and Huron's typology \subfiglabel{(B)}, for two datasets: Erk (blue) and Han (orange).
The theoretical shape is shown in the background (see Figure~1 in the main text).
For the $k$-means typology \subfiglabel{(C)} we show the centroids for $k=3$, $4$ and $5$ clusters.
Similarly shaped centroids are similarly colored across values of $k$.
The shapes in smaller typologies ($k$-means or Huron's) are more recognizable than those in Adams' typology.
Also note the characteristic flattening at the beginning and end of each contour, caused by every first and final note necessarily being flat.

\end{document}